# Incorporation of Non-metal Impurities at the Anatase TiO$_2$(001)-(1×4) Surface

Jun Hee Lee,[1*] Daniel Fernandez Hevia,[2] and Annabella Selloni[1]

[1]Department of Chemistry, Princeton University, Princeton, New Jersey 08544, USA

[2] Universidad de Las Palmas de Gran Canaria, Campus de Tafira, 35017 Las Palmas de Gran Canaria, Spain

**Abstract**

We use first-principles calculations to investigate the adsorption and incorporation of non-metal impurities (N, C) at the anatase TiO$_2$(001)-(1×4) reconstructed surface. We analyze in detail the influence of the surface structure and local strain on the impurity binding sites and incorporation pathways and identify important intermediates which facilitate impurity incorporation. We find various subsurface interstitial binding sites and corresponding surface → subsurface penetration pathways on the reconstructed surface. This surface also favors the presence of subsurface oxygen-vacancies, to which adsorbed species can migrate to form substitutional impurities. Most notably, we show that the non-exposed oxygen sites just below the surface have a key role in the incorporation of nitrogen and carbon in TiO$_2$(001).

[*] Corresponding author: junhee@princeton.edu

Titanium dioxide (TiO$_2$) is widely used in photocatalysis [1,2,3,4], and solar energy conversion[5,6,7], yet it is not very efficient. One of the most serious drawbacks of TiO$_2$ is its large band gap, $E_g \sim$ 3.2 eV, which results in the absorption of only a small portion of the solar spectrum in the UV region. An essential prerequisite for increasing the photocatalytic efficiency of TiO$_2$ is to reduce its band gap. Asahi et al.[8] first suggested that this can be achieved using anion dopants, particularly nitrogen, to replace lattice oxygens. The huge amount of experimental and theoretical studies which followed has clarified the electronic structure of anion-doped TiO$_2$ to a large extent[9,10,11,12]. Still, important questions remain, such as the factors which limit the impurity incorporation in TiO$_2$ and the influence of doping on the overall photochemical activity[5,13].

The physico-chemical properties of specific surfaces also play a key role in photocatalysis[14,15]. Here we focus on the anatase TiO$_2$(001) surface which was predicted to be very reactive on the basis of first-principles calculations[16,17], and has thus attracted considerable interest in recent years. Experimentally, the (001) surface is found to exhibit a (1×4) reconstruction which is stable in a wide temperature range and under a variety of conditions[18,19]. This reconstruction is present also on the surface of N-doped samples[13,20], which is interesting because of two inter-related effects. On the one hand, studies on silicon and other semiconductors have shown that the surface structure and intrinsic stress have an important influence on the incorporation and spatial distribution of impurities[21], suggesting that similar effects may take place also at the TiO$_2$(001) surface. On the other hand, surface and subsurface impurities are expected to substantially affect the photochemical properties, implying that knowledge of the impurity distribution in the surface region is important for understanding the influence of doping on the photocatalytic activity[22]. Despite the widespread interest in the anatase (001) surface, its capability to accommodate impurities has not yet been explored.

In this work we use first-principles density functional theory (DFT) calculations to investigate the adsorption and incorporation of non-metal impurities at the anatase TiO$_2$(001) surface. We focus on nitrogen, which has attracted much attention in recent years, but, for comparison, we consider also carbon, another widely used *p*-type dopant[23]. We examine the character and energetic of adsorption vs interstitial and substitutional doping at the (1×4)-reconstructed surface, comparing to results for the bulk and analyzing the role of the surface structure in the stability of different species. Starting from adsorbed species, we then study likely incorporation pathways for both interstitial and substitutional impurities and identify key intermediates which can lead to facile impurity incorporation in TiO$_2$(001).

Our study is based on spin-polarized DFT calculations in the Generalized Gradient Approximation (GGA) of Perdew-Burke-Ernzerhof (PBE)[24] and the plane-wave-pseudopotential scheme as implemented in the PWSCF code of the Quantum Espresso package[25]. Additional calculation details are given in Supplement I, while comparative results of selected calculations using the DFT+U scheme[26] are reported in Supplement II. To describe the reconstructed $TiO_2$(001)-(1×4) surface, we use the "ad molecule" (ADM) model[27], which agrees well with the available experimental information and is widely accepted. This model is characterized by added $TiO_2$ rows forming "ridges" which run parallel to the [010] direction and expose four-fold coordinated Ti atoms ($Ti_{4c}$). In-between ridges, "terraces" expose five-fold coordinated Ti ($Ti_{5c}$) and two-fold coordinated O ($O_{2c}$) atoms, and have the structure of the unreconstructed surface with a lateral compression along [100]. It is mainly the presence of this lateral compression that strongly stabilizes the reconstructed (001) surface by relieving the large intrinsic tensile stress of the unreconstructed surface[27].

Figure 1 shows the formation energies and several geometries of adsorbed and incorporated nitrogen impurities at the $TiO_2$(001)-(1×4) surface. Due to the presence of many inequivalent sites at the reconstructed surface, a variety of species can be identified. In particular, $N^a$ and $N^b$ denote two different adsorption configurations on the terraces, with N-Ti bonds in (010) and (100) planes, respectively; similarly, $N^a_R$ ($N^b_R$) indicates $N^a$ ($N^b$)-type adsorption in the ridge (R) region. From Figure 1, we can see that these adsorbed species have formation energies similar to those of some subsurface interstitials ($N_{i1}$-$N_{i4}$), indicating that subsurface interstitial incorporation is competitive with adsorption on the reconstructed surface. We investigated the effect of the surface reconstruction on the structure and stability of the different species by performing calculations for nitrogen at the unreconstructed $TiO_2$(001) surface. We found that some structures exist only on the reconstructed surface, notably $N^a$ bonded to a single Ti atom (upper left of Figure 1). In addition subsurface interstitials are significantly more stable at the reconstructed surface. The influence of the reconstruction is evident in the case of the $N_{i4}$ subsurface interstitial (bottom right of Figure 1): due to the local compressive strain in the terraces, the nitrogen interstitial pushes the nearest oxygen ($O_{2'}$) outside the surface and takes its original lattice position, effectively becoming a $N_{O2'}$ substitutional impurity capped by a surface oxygen. The resulting $N_{i4}$ species is energetically more stable than a bulk interstitial, $N_{i\ bulk}$, by more than 1.1 eV.

The reconstructed surface provides also favorable pathways for N adatoms to migrate below the surface as interstitials. For instance, the $N^b_R$ adsorption configuration transforms spontaneously to a subsurface split-interstitial, $N_{i2}$ (top right of Figure 1). The instability of $N^b_R$ can be attributed to the fact that the surface reconstruction opens a large space under the ridge around $O_2$, which can thus move easily to accomodate the adsorbed nitrogen into the interstitial site. Another migration pathway, from $N^a$ to the "substitutional-like" $N_{i4}$ interstitial, is shown in Figure 2. This pathway involves a $N^b$ intermediate, as no direct $N^a \rightarrow N_{i4}$ pathway was found. Interestingly, the energy barrier for the transformation of $N^b$ to $N_{i4}$, is rather low, ~0.6 eV, suggesting that direct adsorption of nitrogen atoms in the $N^b$ configuration would lead to facile incorporation of N impurities. Although $N^b$ adsorption is energetically less stable than $N^a$ (Figure 1), possibly experimental procedures such as sputtering could increase the population of $N^b$ species, which will subsequently transform to $N_{i4}$ subsurface species. It should also be noted that while interstitials are usually detrimental for the photocatalytic activity because they enhance the recombination rate of the photogenerated carriers, this should not be the case for $N_{i4}$, which has the structural characteristics of a substitutional impurity rather than an interstitial.

Unlike adsorbed species and interstitials, the formation energies of substitutional impurities depend on the value of the oxygen chemical potential ($\mu_O$). Substitutional doping thus becomes the most stable form of nitrogen incorporation for $\mu_O \leq \sim -1.5$ eV (Figure 1), as typically used for the epitaxial growth of anatase $TiO_2$[13,18,19]. At variance with the case of adsorption and interstitial incorporation, the formation energies of substitutional impurities in the terraces ($N_{Osurf}$) of the reconstructed surface are not significantly different from those in the bulk ($N_{Obulk}$), a result consistent with the experimental finding that the maximum concentration of substitutional nitrogen in anatase $TiO_2$ thin-films is ~ 1%[13], close to that obtained in the bulk[8,28]. However, our calculations also predict that the energetic cost of substitutional doping in the ridge ($N_{O0R}$ and $N_{O1R}$) is significantly lower compared to that in the bulk, suggesting that the doping concentration in very thin films of anatase $TiO_2$ might be made higher than in the bulk.

A plausible mechanism for the formation of anionic substitutional impurities is via migration of adatoms to O-vacancies ($V_O$'s) below the surface. This mechanism should be quite effective in anatase where subsurface $V_O$'s tend to be more stable than surface $V_O$'s[29,30]. We examined this incorporation pathway for $N^a_R$ and $N^a$ adsorbed species migrating to $V_{O2}$ and $V_{O2'}$, respectively (Figure 3), which are favorable sites for oxygen vacancy formation on the reconstructed (001) surface[29]. We found that the migration barrier is almost vanishing for $N^a$, on the terrace, whereas it is significant, ~1.0 eV, for $N^a_R$, on the ridge. For the latter the migration requires indeed the breaking of a bond with a Ti on the ridge, whereas $N^a$ can smoothly move to the subsurface to form a substitutional $N_{O2'}$ without breaking any Ti-N bond. It is interesting to remark the close similarity between this $N_{O2'}$ species and the "substitutional-like" $N_{i4}$, which, as mentioned earlier, is just $N_{O2'}$ capped by a surface oxygen. Both $N_{i4}$ and $N_{O2'}$ are key intermediates for N incorporation at the $TiO_2$(001) surface. Their structural similarity suggests that direct interconversion between them can occur, depending on the value of $\mu_O$. As a result, nitrogen incorporation in TiO$_2$(001) is practically controlled by a single species, N$_{O2'}$/N$_{i4}$, involving the subsurface O$_{2'}$ sites on the terraces.

Carbon is another frequently used *p-type* dopant in TiO$_2$[23]. As shown in Figure 4a, for carbon the energetic cost of substitutional incorporation at the TiO$_2$(001)-(1×4) surface is quite high so that, different from nitrogen, substitution is generally unfavorable relative to interstitial doping. However, the most stable C interstitial is C$_{i4}$ (Figure 4b), which is analogous to the N$_{i4}$ species discussed above. Also carbon's incorporation pathways, from adatoms to subsurface interstitials, are similar to those of nitrogen. There is a pathway from C$^b_R$ to C$_{i2}$ analogous to the N$^b_R$ → N$_{i2}$ transformation in Figure 1; the only difference is that, instead of being barrierless as for nitrogen, for carbon there is a barrier of ~ 0.4 eV (see Supplement III). Analogous to the N$^b$ → N$_{i4}$ pathway in Figure 2, there is also a path bringing from adsorbed C$^b$ to the "substitutional-like" C$_{i4}$, see Figure 4b; although the barrier is quite high, this should still provide a convenient pathway for carbon incorporation at high temperature. Altogether, our calculations indicate that the surface reconstruction affects the adsorption and incorporation of *p*-type dopants nitrogen and carbon in similar ways. In particular, similar to the role of N$_{i4}$ in the case of nitrogen doping, the C$_{i4}$ species is key to carbon incorporation in TiO$_2$(001).

Once the impurities have been incorporated in the subsurface, they can diffuse toward the bulk. To characterize this process, we assume that diffusion barriers below the second layer are essentially the same as in the bulk. The energy barriers of N and C migration through an oxygen vacancy in the bulk are compared to the oxygen vacancy self-diffusion in Figure 5. For completeness, also the diffusion of fluorine, a *n-type* non-metal impurity, is considered. The results show clear trends: the diffusion barriers are large for N and C, they are very small for F, while oxygen is intermediate. These trends can be simply understood considering that N and C tend to form multiple bonds, whereas F prefers to be singly coordinated. Figure 5 shows also a significant anisotropy of the impurity diffusion. Anatase TiO$_2$ can be considered a layered structure along [001], and the atomic in-plane density is higher than that out-of-plane. This gives rise to a large anisotropy in Young's modulus and elastic constants, resulting in anatase being stiffer in the xy plane than along the z-direction[31]. Moreover, as shown in Figure 5, an impurity has to break two bonds with Ti atoms in order to migrate along [110] while only one bond needs to be broken for interlayer diffusion along [301]. This explains why out-of-plane diffusion across adjacent TiO$_2$ layers is significantly more facile than diffusion within the TiO$_2$ layer for all the investigated impurities.

By combining the results in Figures 3 and 5, we can see that after a N adatom migrates to the oxygen O$_{2'}$ substitutional site, penetration into the bulk by hopping across adjacent TiO$_2$ layers has a barrier of 0.63 eV (0.73 eV using DFT+U). This is larger than the corresponding oxygen self-diffusion barrier of 0.15 eV (0.39 eV using DFT+U), suggesting a poor mobility of N impurities at room temperature and below. On the other hand, the fact that the oxygen self-diffusion barrier is quite small suggests that electron irradiation of anatase $TiO_2$ (001) could provide an effective way to create and propagate oxygen vacancies inside the bulk region. This could in turn help N substitutional incorporation into the bulk.

In conclusion, we have presented first-principles calculations which provide insights into the mechanisms of incorporation of N and C, two widely used non-metal impurities, at the reconstructed anatase $TiO_2(001)$-$(1\times 4)$ surface. Our results point to the central role of surface structure and local stress in determining preferential adsorption and incorporation at selected sites as well as the existence of favorable incorporation pathways and intermediates. In particular, our results show that the subsurface $O_{2'}$ sites in the terraces play a crucial role in the incorporation of non-metal impurities. The ability to access these sites provides the key to control and/or enhance the incorporation and photochemical activity of anion dopants in $TiO_2(001)$.

This research has been supported by Spanish MICINN through Program INNPACTO Project CASCADA IPT-120000-2010-19. We also acknowledge use of the TIGRESS high performance computer center at Princeton University.

**FIGURE CAPTIONS**

**Figure 1.** Formation energies (eV) vs. oxygen chemical potential ($\mu_O$) or $O_2$ pressure at fixed temperatures ($T$ = 300 K and 900 K, top) for N impurities at the anatase $TiO_2$(001)-(1×4) surface. The range of $\mu_O$ typically used in experiment is highlighted in yellow. For comparison, the formation energies of substitutional ($N_{O,bulk}$) and interstitial ($N_{i,bulk}$) impurities in bulk anatase are also reported. On the left and right sides of the phase diagram, inequivalent adsorption and subsurface interstitial structures are shown. Numbers 1~4'' indicate inequivalent oxygen sites. Large yellow and small blue spheres represent oxygen and titanium atoms, respectively; adsorbed nitrogen species are black and subsurface nitrogen interstitials are red. The pink arrow in the top right structure indicates the spontaneous relaxation from $N^b_R$ to $N_{i2}$. Directions **a**, **b**, **c** correspond to [100], [010], and [001], respectively.

**Figure 2.** Minimum energy pathway (MEP) for the diffusion of a N impurity from adsorption $N^a$ to interstitial $N_{i4}$ through a $N^b$ adsorption intermediate. Atomic structures of selected configurations along the MEP are shown. Integers represent the different images in the NEB calculation. Bold (dotted) arrows denote stable (transition) states.

**Figure 3.** MEP for the diffusion of N impurities from the adsorption $N^a_R$ and $N^a$ sites to the substitutional $O_2$ (top) and $O_{2'}$ (bottom) sites, respectively. Atomic structures of selected configurations along the MEP are shown on the bottom. Integers represent the different images in the NEB calculations. Blue (red) dotted arrows denote transition states from $N^a_R$ ($N^a$).

**Figure 4.** (a) Formation energies (eV) vs. oxygen chemical potential or $O_2$ pressure at fixed temperatures ($T$ = 300 K and 900 K, top) for C impurities at the anatase $TiO_2$(001)-(1×4) surface. For comparison, the formation energies of substitutional ($C_{O,bulk}$) and interstitial ($C_{i,bulk}$) impurities in bulk anatase are also reported. Labels of different species are the same as for nitrogen in Figure 1. (b) MEP incorporation pathway from adsorption $C^b$ to interstitial $C_{i4}$. The geometries of these states are also shown.

**Figure 5.** Diffusion pathways and corresponding potential energy barriers for C, N, F impurities and for an oxygen vacancy in bulk anatase. Atomic structures of initial and final configurations along the MEP are shown in the upper part of the figure.

**Reference**

[1] A. L. Linsebigler, G. Lu, and J. T. Yates, Chem. Rev. **95**, 735 (1995).

[2] M. R. Hoffmann, S. T. Martin, W. Choi, and D. W. Bahnemann, Chem. Rev. **95**, 69 (1995).

[3] T. L. Thomson and J. T. Yates, Chem. Rev. **106**, 4428 (2006).

[4] M. A. Henderson, Surf. Sci. Rep. **66**, 185 (2011).

[5] M. Grätzel, Nature **414**, 338 (2001).

[6] A. Fujishima, X. Zhang, and D. A. Tryk, Surf. Sci. Rep. **63**, 515 (2008).

[7] X. Chen and S. S. Mao, Chem. Rev. **107**, 2891 (2007).

[8] R. Asahi, T. Morikawa, T. Ohwaki, K. Aoki, and Y. Taga, Science **293**, 269 (2001).

[9] Y. Gai, J. Li, S.-S. Li, J.-B. Xia, and S.-H. Wei, Phys. Rev. Lett. **102**, 036402 (2009).

[10] P. Wang, Z. Liu, F. Lin, G. Zhou, J. Wu, W. Duan, B.-L. Gu, and S. B. Zhang, Phys. Rev. B **82**, 193103 (2010).

[11] C. Di Valentin and G. Pacchioni, Catal. Today (2011), in press.

[12] J. B. Varley, A. Janotti, and C. G. Van de Walle, Adv. Mater. 23, 2343 (2011).

[13] T. Ohsawa, I. Lyubinetsky, Y. Du, M. A. Henderson, V. Shutthanandan, and S. A. Chambers, Phys. Rev. B **79**, 085401 (2009).

[14] U. Diebold, Surf. Sci. Rep. **48**, 53 (2003).

[15] T. Ohno, K. Sarukawa, and M. Matsumura, New J. Chem. **26**, 1167 (2002).

[16] A. Vittadini, A. Selloni, F. P. Rotzinger, and M. Grätzel, Phys. Rev. Lett. **81**, 2954 (1998).

[17] X. Q. Gong and A. Selloni, J. Phys. Chem. B **109**, 19560 (2005).

[18] G. S. Herman, M. R. Sievers, and Y. Gao, Phys. Rev. Lett. **84**, 3354 (2000).

[19] Y. Liang, S. Gan, S. A. Chambers, and E. I. Altman, Phys. Rev. B **63**, 235402 (2001).

[20] S. H. Cheung, P. Nachimuthu, M. H. Engelhard, C. M. Wang, and S. A. Chambers, Surf. Sci. **602**, 133 (2008).

[21] J. Tersoff, Phys. Rev. Lett. **74**, 5080 (1995).

[22] G. Liu, L. Wang, H. G. Yang, H.-M. Cheng, and G. Q. Lu, J. Mat. Chem. **20**, 831 (2010).

[23] S. Sakthivel and H. Kisch, Angew. Chem., Int. Ed. **42**, 4908 (2003).

[24] J. P. Perdew, K. Burke, and M. Ernzerhof, Phys. Rev. Lett. **77**, 3865 (1996).

[25] S. Baroni, P. Giannozzi, S. De Gironcoli, and A. Dal Corso, QUANTUM ESPRESSO, http://www.democritos.it.

[26] V. I. Anisimov, J. Zaanen, and O. K. Anderson, Phys. Rev.B **44**, 943 (1991).

[27] M. Lazzeri and A. Selloni, Phys. Rev. Lett. **87**, 266105 (2001).

[28] H. Irie, Y. Watanabe, and K. Hashimoto, J. Phys. Chem. B **107**, 5483 (2003).

[29] H. Cheng and A. Selloni, Phys. Rev. B **79**, 092101 (2009).

[30] We notice that formation of subsurface oxygen vacancies is very unfavorable on the unreconstructed $TiO_2$(001) surface, due to its large tensile strain. Therefore the substitutional incorporation impurities in the subsurface region is hindered on the unreconstructed surface.

[31] W.-J. Yin, S. Chen, J.-H. Yang, X.-G. Gong, Y. Yan, and S.-H. Wei, Appl. Phys. Lett. **96**, 221901 (2010).

# Figure 1

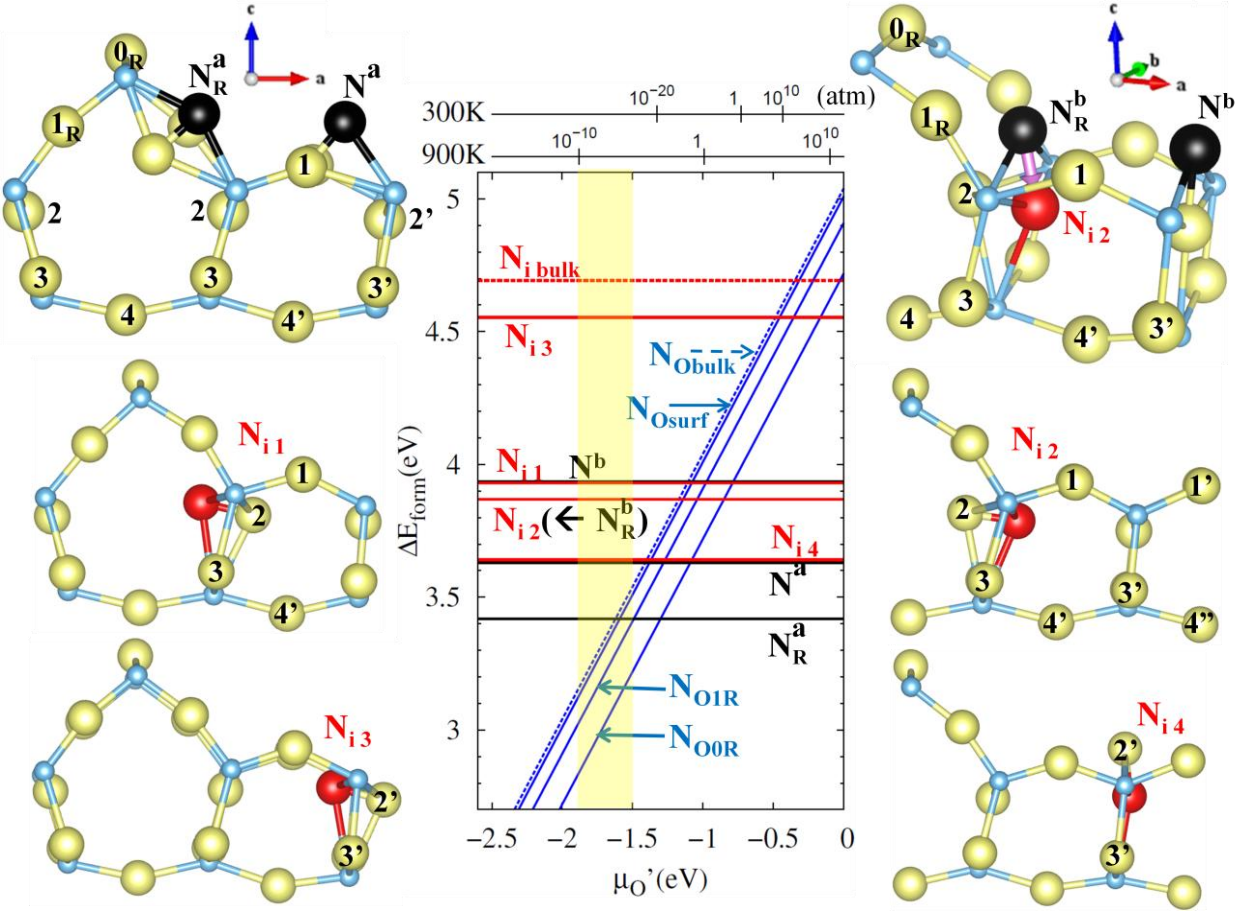

**Figure 2**

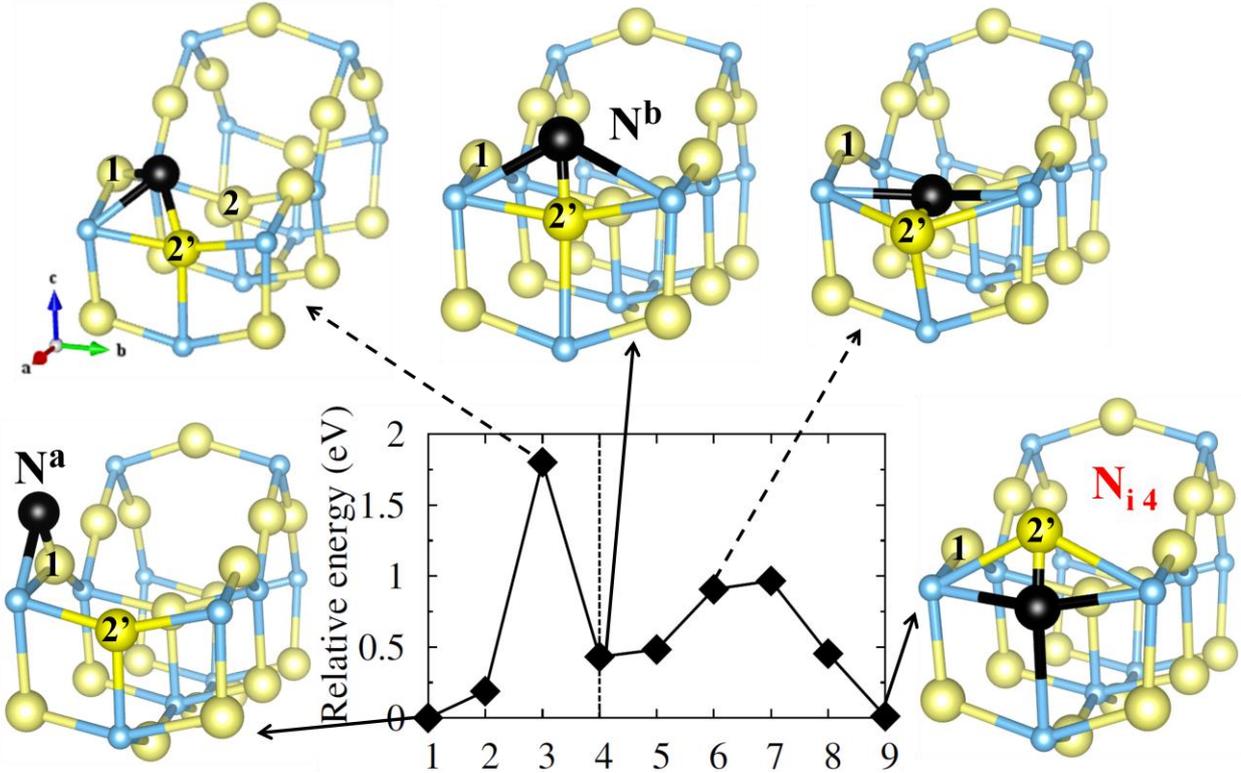

**Figure 3**

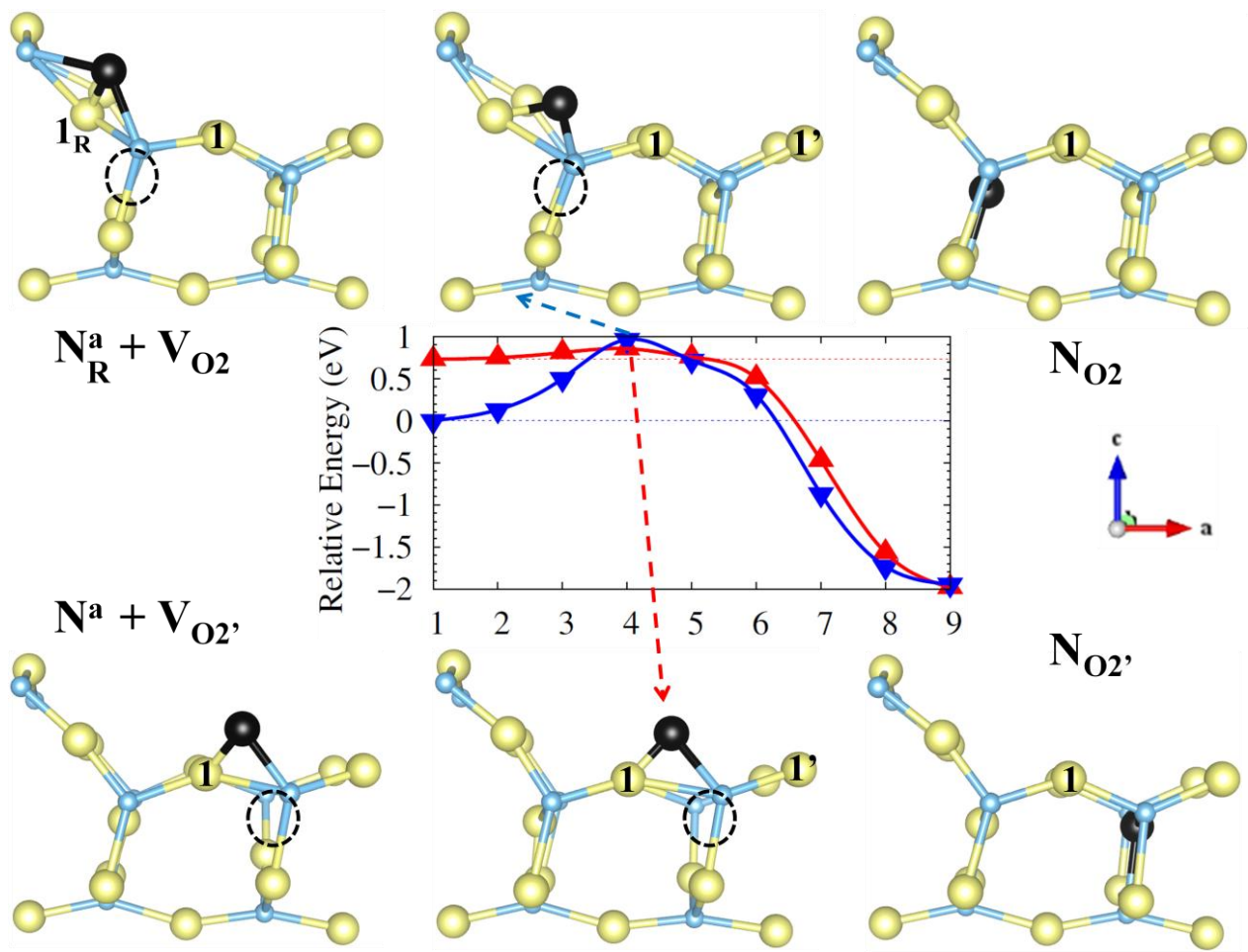

**Figure 4**

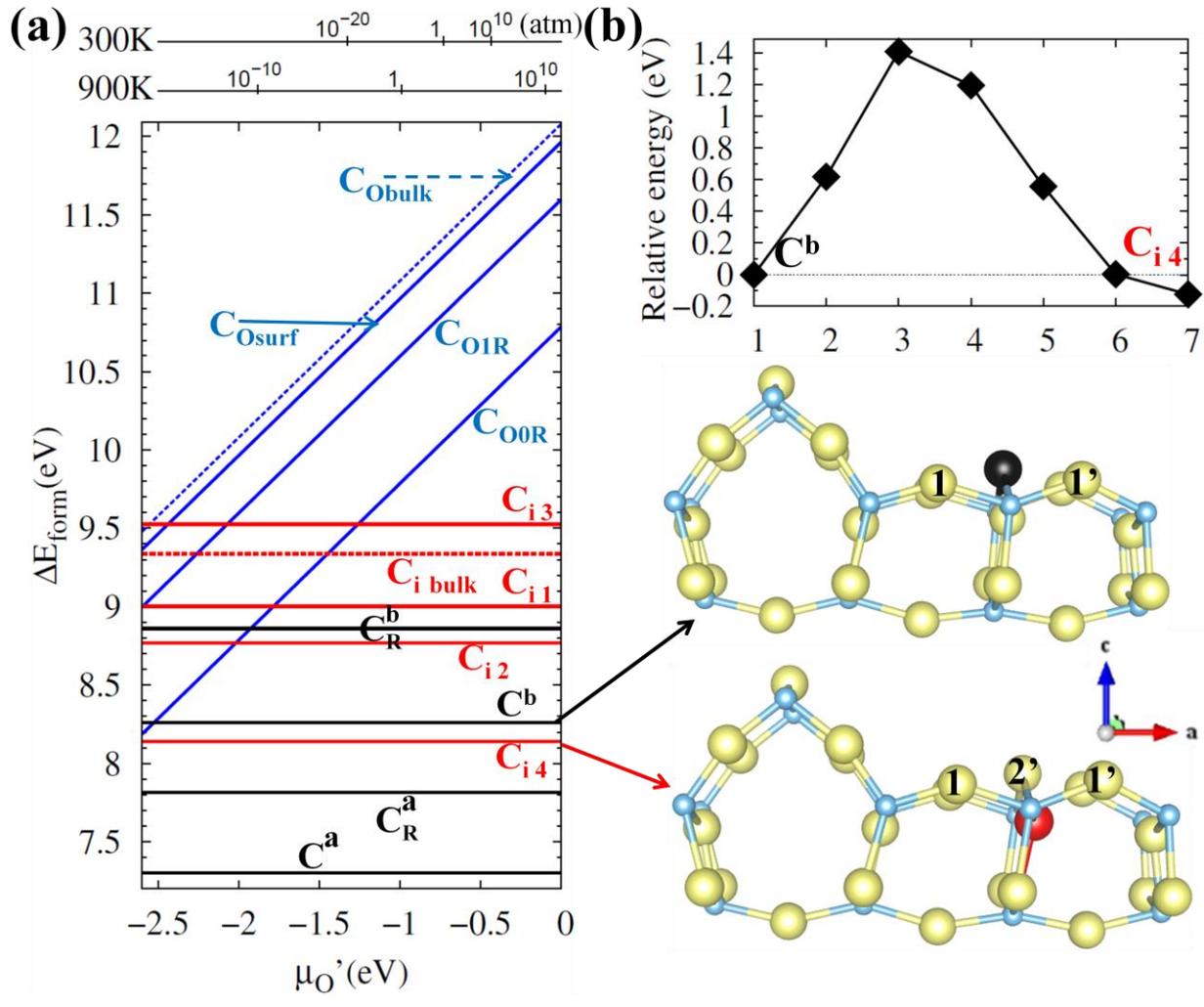

**Figure 5**

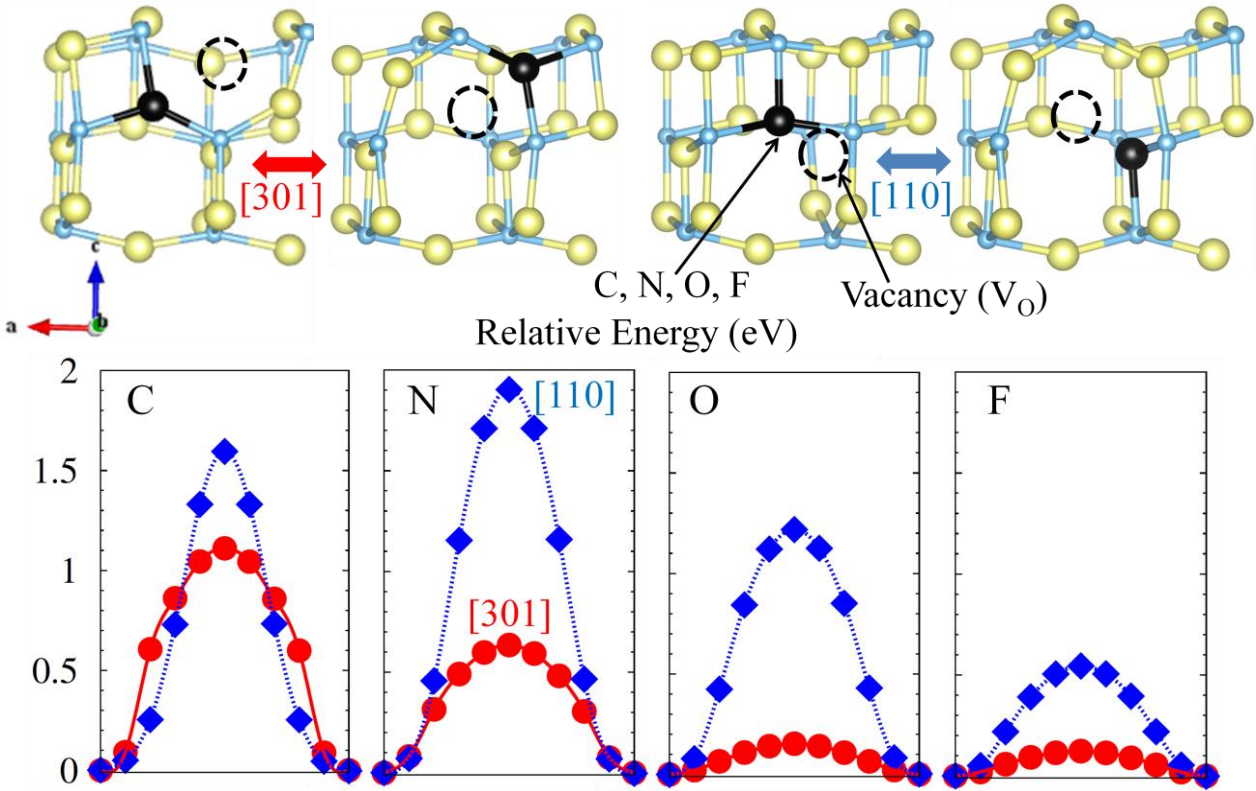

**Supplement I – Calculation Method**

Our study is based on spin-polarized DFT calculations in the Generalized Gradient Approximation (GGA) of Perdew-Burke-Ernzerhof (PBE)[31] and the plane-wave-pseudopotential scheme[31] as implemented in the PWSCF code of the Quantum Espresso package.[31] In the case of F-doping, preliminary test calculations showed the influence of spin polarization to be negligible; therefore subsequent calculations did not include the spin polarization. We also performed test calculations using the GGA+$U$ method[31] (with $U = 3.5$ eV [31] on the Ti 3d states) to determine the formation energies of several substitutional sites; as reported in Supplement II, the differences between GGA and GGA+$U$ results are extremely small. Plane-wave basis sets cutoffs were set at 30 and 240 Ry for the smooth part of the wave function and the augmented charge density, respectively.

To model the (1×4) reconstructed surface, we used slabs of five layers with a (2×4) surface supercell and a total of 126-atoms; consecutive slabs were separated by a 11 Å vacuum. The experimental in-plane lattice constant ($a_0 = 3.782$ Å) was used (the calculated equilibrium lattice parameter is only slightly different, $a_0 = 3.786$ Å), and k-space was sampled with a 2×1×1 $k$-point mesh. The bottom $TiO_2$ layer was fixed at its bulk position, and residual forces were smaller than $4 \times 10^{-4}$ au. For comparison of surface and bulk formation energies, we used a 2×4×1 bulk supercell with 96-atoms and 2×1×1 $k$-point mesh, and relaxed the bulk supercell until the residual pressure became smaller than 0.05 Kbar. The nudged elastic band (NEB) method[31] was used to determine diffusion pathways and energy barriers. Bulk diffusion pathways were studied using a 3×3×1 bulk supercell and sampling the Brillouin zone using only Γ; detailed test calculations confirmed that the results did not change significantly with increasing the number of $k$-points.

Substitutional doping was studied by replacing one oxygen per supercell by an impurity atom. This corresponds to $TiO_{2-x}A_x$ ($A$ = N, C) with x = 0.0238. Interstitial doping and adsorption were studied by adding one impurity atom at the appropriate site. Formation energies were determined using the expression: $E_{form} = E_{tot}(A\text{-doped}) - E_{tot}(\text{pure}) - \mu(A) + \mu(O)$ for substitution, and $E_{form} = E_{tot}(A\text{-doped}) - E_{tot}(\text{pure}) - \mu(A)$ for adsorption and interstitial doping, where μ(A) denotes the chemical potential of species $A$. We used fixed values of the chemical potential, $\mu(N) = [1/2]\mu(N_2)$ for N, and $\mu(C) = \mu(CO_2) - \mu(O_2)$ for C, while allowing the oxygen chemical potential to vary according to $\mu(O) = [1/2]\mu(O_2) + \mu'(O)$; $\mu(N_2)$, $\mu(O_2)$, and $\mu(CO_2)$ are the calculated total energies of the $N_2$, $O_2$ and $CO_2$ molecule, respectively. The value of the oxygen chemical potential was also converted to unit of oxygen pressure at fixed temperatures of 300 K and 900 K (typical annealing temperature for N-doped anatase $TiO_2$). We used VESTA[31] for visualization of the calculated geometries.

**Supplement II.** Formation energies (in eV) of N- and C-substitutional impurities at different sites of the anatase $TiO_2$(001)-(1×4) surface, computed at the GGA (GGA+$U$) level. Oxygen sites are denoted as in Figures 1.

|  | Site # | Nitrogen | N-avg. | Carbon | C-avg. |
|---|---|---|---|---|---|
| Ridge | $O_{0R}$ | 4.72 (4.70) | 4.82 | 10.79 | 11.20 |
|  | $O_{1R}$ | 4.91 (4.91) |  | 11.60 |  |
| Terrace First layer | $O_1$ | 5.03 | 5.01 | 11.69 | 11.97 |
|  | $O_{1'}$ | 4.97 (4.91) |  | 11.66 |  |
|  | $O_2$ | 5.11 |  | 12.02 |  |
|  | $O_{2'}$ | 5.06 |  | 12.20 |  |
| Terrace Second layer | $O_3$ | 4.97 (4.95) |  | 11.88 |  |
|  | $O_{3'}$ | 4.97 (4.95) |  | 12.24 |  |
|  | $O_4$ | 4.89 |  | 11.83 |  |
|  | $O_{4'}$ | 5.02 |  | 11.94 |  |
|  | $O_{4''}$ | 5.12 |  | 12.23 |  |
| Bulk | O | 5.04 | 5.04 | 12.08 | 12.08 |

**Supplement III.** Minimum energy pathway (MEP) for the diffusion of a C adatom from $C^b_R$ to the interstitial $C_{i2}$ site on anatase $TiO_2$(001)-(1×4) surface. Atomic structures of selected configurations along the MEP are shown. Integers indicate the different images in the NEB calculation. Bold (dotted) arrows denote stable (transition) states.

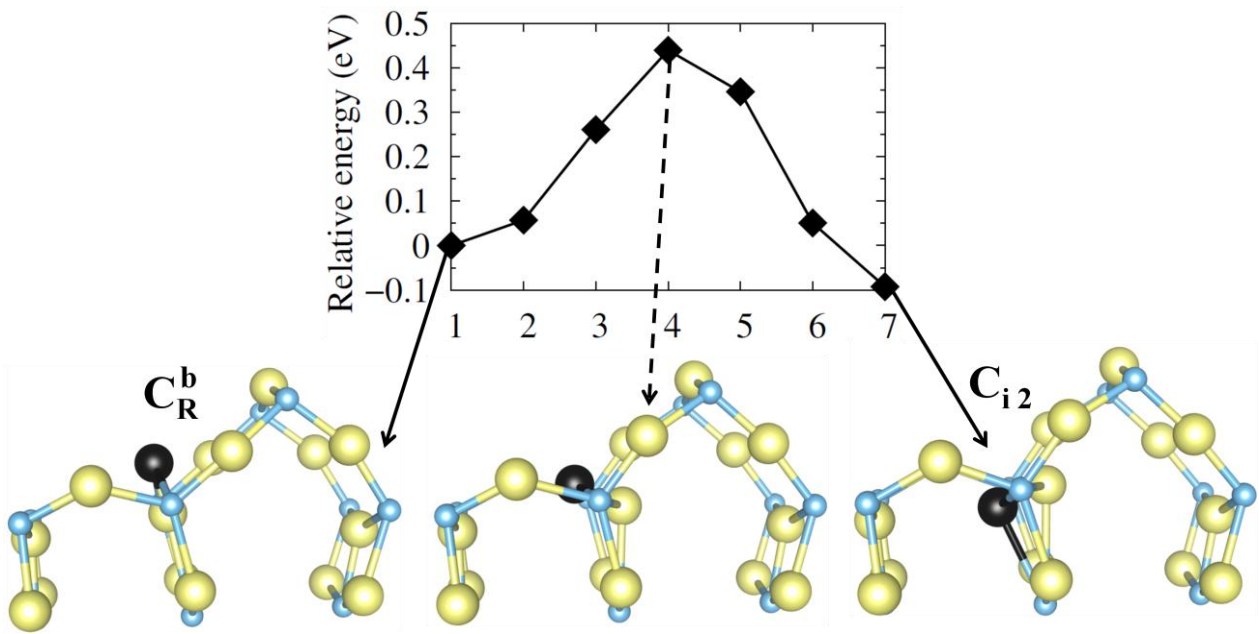